\documentstyle[aps,twocolumn,epsf,prb]{revtex}
\begin{document}
\draft
\title{Tunneling magnetoresistance in diluted magnetic
semiconductor tunnel junctions}
\author{Pin Lyu \cite{Lyu} and Kyungsun Moon}
\address{Department of Physics and Institute of Physics and Applied Physics,
Yonsei University, Seoul 120-749, Korea}
\date{\today}
\maketitle
\begin{abstract}
Using the spin-polarized tunneling model and taking
into account the basic physics of ferromagnetic
semiconductors, we study the temperature dependence of
the tunneling magnetoresistance (TMR) in
the diluted magnetic semiconductor (DMS) trilayer heterostructure system
(Ga,Mn)As/AlAs/(Ga,Mn)As. The experimentally observed TMR ratio is
in reasonable agreement
with our result based on the typical material parameters.
It is also shown that the TMR ratio has a strong dependence on
both the itinerant-carrier density and the magnetic ion density
in the DMS electrodes.
This can provide a potential way to achieve larger TMR ratio
by optimally adjusting the material parameters.

\vskip 0.2in


\end{abstract}

\section{INTRODUCTION}
Since the early experiment done by Julliere,\cite{Ju} there has been considerable
interest in studying the tunneling magnetoresistance (TMR) effect 
in magnetic tunnel junctions
\cite{s1,s2}
due to its promising applications in digital storage and magnetic
sensor technology.\cite{s3}
The conventional electrodes of magnetic tunnel junctions are made up
of the  ferromagnetic transition metals such as Fe, Co, Ni,
and their alloys. Usually the TMR effect in the conventional magnetic tunnel
junctions is described by the Julliere model.\cite{Ju,2}
The electrons near at the Fermi levels of the electrodes participate in the
tunneling process and have the main contribution to the tunneling 
conductance and so to the TMR effect.

Recent discovery of the ferromagnetism in the diluted magnetic semiconductors (DMS)
has opened the rich field of an interesting interplay between magnetic and semiconducting
properties,\cite{ohno,4} which makes it possible to combine the information 
processing and data storage in one material.
At high concentration $c$ of randomly distributed ${\rm Mn}^{2+}$ ions doped 
in GaAs samples with high hole density
$c^{*}$,  Ga$_{1-x}$Mn$_{x}$As compounds exhibit ferromagnetism with
transition
temperature as high as 110 K at certain value of $x$. The origin of
ferromagnetism in the DMS can be explained by the itinerant holes through the 
$p$-$d$ exchange coupling
between the valance-band holes and the spins of impurity Mn 
ions.\cite{ohno,4,Mac}
In order to investigate the spin-polarized electron transport
phenomenon in the DMS ferromagnetic semiconductor heterostructures,
the epitaxial tunnel junctions have been successfully fabricated in the form of
Ga$_{0.961}$Mn$_{0.039}$As/AlAs/Ga$_{0.961}$Mn$_{0.039}$As.
The TMR effect was observed and the maximum TMR ratio thus obtained is as high 
as 44.4\% at 4.2 K.\cite{1} The subsequent
experiment \cite{apl} also realized the TMR effect in the DMS tunnel junctions of
Ga$_{0.95}$Mn$_{0.05}$As/Al$_{y}$Ga$_{1-y}$As/Ga$_{0.97}$Mn$_{0.03}$As.

The Julliere model \cite{Ju,2} with several necessary extensions
\cite{JCS,Gu,5,Lyu1,Lyu2}
is known to describe the tunnel
junctions, in which the two electrodes are made of the conventional 
ferromagnetic metals or
the {\em half-metallic} manganites such as La$_{1-x}$Sr$_{x}$MnO$_{3}$.
For the manganite tunnel junctions, various tunneling processes are included
in the model as follows. The {\em elastic} spin-flip scattering of the spin 
polarized
carriers by the local magnetic impurities within the insulating barrier and 
the {\em inelastic} spin-flip scattering induced by the thermal collective
excitations of local spins at the interfaces between the insulating barrier
and the manganite electrodes are the two basic ingredients.
By taking into account the temperature dependence of the spin polarization of
itinerant carriers, one can calculate the temperature dependence of the TMR ratio.\cite{Lyu2}
In the paper, we have applied the above generalized Julliere model to the DMS 
tunnel junctions and obtained the various properties of the TMR effects of the system.

The paper is organized as follows.
In Sec.\ II, we propose a spin-polarized tunneling model for the DMS
tunnel junctions  by generalizing the previous spin-polarized tunneling model 
in order to correctly take into account
the basic physics of the DMS. In Sec.\ III, we show that the experimentally
observed TMR ratio in the DMS tunnel
junctions can be well reproduced from our calculations with the appropriate
choice of
typical material parameters and also show that
the TMR ratio thus obtained is strongly dependent
on both the itinerant-carrier density and the magnetic ion density in the DMS
electrodes. Finally we conclude with a brief summary in Sec.IV.

\section{SPIN-POLARIZED TUNNELING MODEL}

We will briefly summarize the basic physics of ferromagnetism occurring in
certain DMS
compounds following the seminal work of K\"{o}nig {\em et al.}.\cite{Mac}

In the doping range of our interest, that is, $0.035 < x < 0.06$, Ga$_{1-x}$Mn$_{x}$As 
can be described as
a degenerate semiconductor with relatively high hole concentrations and the 
${\rm Mn}^{2+}$ ions act as local magnetic impurities.

The Hamiltonian describing the basic physics
of the DMS can be written as

$${\cal H}=\int d^{3}r  \sum  _{\sigma} \Psi _{\sigma}^{\dag}({\bf r})(-\frac{\hbar ^
{2}}{2m^{*}} \nabla^{2}-\mu ^{*})\Psi _{\sigma}({\bf r})
$$
\begin{equation}+J_{pd}\int d^{3}r {\bf S}({\bf r})\cdot
{\bf s}({\bf r}),
 \end{equation}
where ${\bf S}({\bf r})$ stands for the spin density of the local magnetic
impurities, which is
given by $\sum _{I}{\bf S}_{I}\delta ({\bf r}-{\bf R}_{I})$ with
${\bf R}_{I}$ denoting the positions of magnetic impurities.
The field operator $\Psi^{\dag}_{\sigma}({\bf r})$ ($\Psi_{\sigma}({\bf r})$)
is the creation (annihilation) operator for the itinerant hole
at position ${\bf r}$ with spin $\sigma$.
The ${\bf s}({\bf r})$ represents the spin density of itinerant-carriers,
which can be written in terms of field operators as
$\frac{1}{2}\sum _{\sigma \sigma '}
\Psi ^{\dag}_{\sigma}({\bf r}) \vec {\tau}_{\sigma \sigma '}\Psi _{\sigma
'}({\bf r})$,
where $\vec {\tau}$ are the three Pauli spin matrices. The $m^{*}$ and $\mu^{*}$ 
are the effective mass and chemical potential of itinerant carriers, respectively.
The $J_{pd}$ stands for the $p$-$d$ exchange coupling strength between itinerant 
holes and ${\rm Mn}^{2+}$ ion impurity spins.
The recent experiments suggest that the itinerant hole density $c^{*}$ is
much lower than the magnetic ion density $c$ satisfying $c^{*}/c \ll 1$.
Here we focus our attention on the DMS with ferromagnetic ground state.
Using the Holstein-Primakoff transformations and performing the coarse graining 
procedures, the local spin density ${\bf S}({\bf r})$ can be written in terms 
of the newly defined bosonic fields
$b^{\dag}({\bf r}), b({\bf r})$ as follows
\begin{equation}
S^{+}({\bf r})=\sqrt{2cS-b^{\dag}({\bf r})b({\bf r})}\, b({\bf r}),
\end{equation}
\begin{equation}
S^{-}({\bf r})=b^{\dag}({\bf r}) \, \sqrt{2cS-b^{\dag}({\bf r})b({\bf r})},
\end{equation}
\begin{equation}
S^{z}({\bf r})=cS-b^{\dag}({\bf r})b({\bf r}).
\end{equation}
The self-consistent spin wave approximation developed in Ref.\ \onlinecite{Mac} 
yields the following {\em effective} Hamiltonian for the DMS system
\begin{equation}
{\cal H}_{\rm eff}=\sum _{{\bf k}\sigma} \varepsilon_{\bf k \sigma}c^{\dag}_{{\bf
k}\sigma}c_{{\bf k}\sigma}
+\sum _{|{\bf q}|<q_{c}} \Omega _{\bf q}b^{\dag}_{\bf q}b_{\bf q},
\end{equation}
where $\varepsilon_{\bf k \sigma}$ stands for the band energy of the itinerant 
holes with spin $\sigma = \pm 1$, which is given by
\begin{equation}
\varepsilon_{\bf k \sigma}=\varepsilon_{\bf k}+\frac{\Delta}{2}\sigma
\end{equation}
and  $\varepsilon_{\bf k} = \hbar^{2}k^{2}/(2m^{*})$.
The spin wave dispersion $\Omega_{\bf q}$ is given by
$$\Omega _{\bf q}=\frac{J_{pd}}{2}(n_{\downarrow}-n_{\uparrow})-
\frac{J_{pd}\Delta}{2 V}\sum _{\bf k} \Bigl [ f(\varepsilon_{{\bf k}\downarrow} )
-f(\varepsilon_{{\bf k+q}\uparrow})\Bigl ]
$$
\begin{equation}
\times \frac{1}{\Omega _{\bf q}+\varepsilon
_{{\bf k+q}\uparrow}-
\varepsilon_{{\bf k}\downarrow}}, 
\end{equation}
where $c_{\bf k}$ and $b_{\bf q}$ are the itinerant hole and the spin wave annihilation 
operators in the momentum space
respectively, and $f(x)$ is the Fermi-Dirac distribution function.
Here $n_{\downarrow}$ ($n_{\uparrow}$)
is the spin-down (spin-up) itinerant hole density satisfying
$n_{\uparrow}+n_{\downarrow}=c^{*}$, the exchange gap $\Delta = cJ_{pd}S$, 
and $q_{c}$ is the Debye cutoff for the spin waves of local impurity spins 
with $q_{c}^{3}=6\pi^{2} c$.
At finite temperatures, the spin wave dispersion can be generalized by
imposing the following self-consistency conditions for the finite temperature
exchange gap : $\Delta (T)=J_{pd}\langle S^{z} \rangle$.
Here $\langle S^{z} \rangle$ represents the thermal average of the 
impurity Mn ion spins, which are approximately calculated by the following formula
$$\langle{S^{z}}\rangle=\frac{1}{V}\sum _{|{\bf
q}|<q_{c}}S{\cal B}_s (\beta S\Omega _{\bf q})
$$
\begin{equation}
=\frac{1}{V}\sum _{|{\bf q}|<q_{c}} \bigl \{S-n_B (\Omega _{\bf q})
+(2S+1)n_B [(2S+1)\Omega
_{\bf q}]\bigl \},
\end{equation}
where ${\cal B}_s (x)$ is the Brillouin function,
$n_B (x)$ the Bose-Einstein distribution function, and $\beta =1/k_{B}T$.
The second term in the second equality of Eq. (8) describes how the thermally
induced spin waves from the independent Bose statistics with {\em no limit}
in the number of spin waves indeed reduce the magnetization of the system.
The third term takes into account the correct spin kinematics, which rules 
out the unphysical states.

Based on the above scheme, one can calculate the magnetization of the Mn 
impurity spins and the band structure of the itinerant holes approximately
self-consistently
by solving a set of coupled equations. In this theoretical framework,
the {\em non-monotonic} dependence of the Curie temperature on the
itinerant hole density $c^{*}$ has been successfully obtained.\cite{Mac}

Now we consider the DMS magnetic tunnel junctions in which both electrodes 
are made of the ferromagnetic DMS
Ga$_{1-x}$Mn$_{x}$As compounds.
The carrier tunneling through a thin AlAs insulating barrier involves 
removing a carrier at one side and creating
it at the other side with spin conserving {\em or} spin-flipping.
The spin-flip tunneling processes can be divided into two different
categories: The elastic spin-flip tunneling arising from the
${\rm Mn}^{2+}$ ion impurity spins located within the insulating barrier and
the inelastic spin-flip
tunneling induced by the thermal collective excitations of the impurity 
spins near at the interfaces between the insulating barrier and the DMS electrodes.

For the collinear
magnetization configurations, the tunneling Hamiltonian of the DMS tunnel
junctions can be written as
\begin{equation}
H=H_{\rm L}+H_{\rm R}+H_{\rm T},
\end{equation}
\begin{equation}
H_{\rm L}=\sum _{{\bf p}\sigma} \varepsilon_{\bf p \sigma}c^{\dag}_{{\bf
p}\sigma}c_{{\bf p}\sigma}
+\sum _{|{\bf q}|<q_{c}} \Omega _{\bf q}b^{\dag}_{{\bf q}L}b_{{\bf q},L},
\end{equation}
\begin{equation}
H_{\rm R}=\sum _{{\bf k}\sigma} \varepsilon_{\bf k \sigma}d^{\dag}_{{\bf
k}\sigma}d_{{\bf k}\sigma}
+\sum _{|{\bf q}|<q_{c}} \Omega _{\bf q}b^{\dag}_{{\bf q}R}b_{{\bf q}R},
\end{equation}
$$
H_{\rm T} = \sum\limits_{{\bf kp}\sigma}
\left(T_{\bf kp}^{0}d^{\dag}_{{\bf k}\sigma}c_{{\bf p}\sigma}
+iT'_{\bf kp}d^{\dag}_{{\bf k}\sigma}c_{{\bf p},-\sigma}
+{\rm H.c.}\right)$$

$$ + \frac {1}{\sqrt{V}} \sum\limits_{{\bf kp}{,|{\bf q}|<q_{c}}}\Bigl \{T''_{{\bf
kpq}}
\bigl [S^{z}_{L}({\bf q})+S^{z}_{R}({\bf q})\bigl ]
\bigl (d^{\dag}_{{\bf k}\uparrow}c_{{\bf p}
\uparrow}-d^{\dag}_{{\bf k}\downarrow}c_{{\bf p}\downarrow}\bigl )
$$
$$+{\rm H.c.}
\Bigl \} 
+ \frac {1}{\sqrt{V}} \sum\limits_{{\bf kp}{,|{\bf q}|<q_{c}}} \Bigl
\{T''_{{\bf kpq}}
\bigl [ {\sqrt{2cS_{L}}}b_{{\bf q}L}
+\sqrt{2cS_{R}} b_{{\bf q}R}\bigl ]
$$

\begin{equation}
\times \bigl (d^{\dag}_{{\bf k}
\downarrow}c_{{\bf p}\uparrow}
+c^{\dag}_{{\bf p}\downarrow}d_{{\bf k}\uparrow}\bigl)
+{\rm H.c.}\Bigl \}. 
\end{equation}

Here $H_{\rm L}$ ($H_{\rm R}$) is the effective Hamiltonian for the {\em bulk} DMS in
the left (right) electrode, which has been derived from the
original Hamiltonian of Eq.\ (1) as explained earlier.
The operator $d_{{\bf k}\sigma}$ $(c_{{\bf p}\sigma})$ is the carrier
annihilation operator with momentum ${\bf k}\, ({\bf p})$ and spin $\sigma$ 
of the
right (left) DMS electrode, and
$S^{z}_{\alpha}({\bf q})=cS_{\alpha}-b^{\dag}_{{\bf q}
\alpha}b_{{\bf q}\alpha}$ with $\alpha=L$, $R$.

The term with coefficient $T'_{\bf kp}$ in $H_{\rm T}$ represents
the spin-flip tunneling arising from
the Mn$^{2+}$ ion impurity spins inside the AlAs insulating barrier.
This type of spin-flip tunneling has been introduced to the tunneling
Hamiltonian for the conventional ferromagnetic tunnel junctions\cite{Gu} and
the manganite tunnel junctions. \cite{Lyu1,Lyu2}
The imaginary $i$ for the spin-flip tunneling matrix elements indicates that 
the elastic spin-flip tunneling
processes are incoherent in any magnetization configurations.
As a generalization of the typical tunneling Hamiltonian \cite{zhang} to the 
DMS tunnel junctions, the last two terms have been included in the tunneling
Hamiltonian of Eq.\ (12), which describe the inelastic spin-flip
tunneling process induced by the thermal spin wave collective
excitations of local spins near at the interfaces between the insulating
barrier and the DMS electrodes.
Since this kind of carrier tunneling involves the emission
or absorption of the local spin collective excitations near the left
or right interfaces, it is an inelastic tunneling process.
The inelastic spin-flip tunneling may occur not only directly at the
interfaces but also within the coherence length of the interfaces, which is
relatively long comparing to the barrier width. It then
allows us to use the {\em bulk} spin wave dispersion relation in the
tunneling Hamiltonian.
The fact that the wave vector ${\bf k}$
is independent of ${\bf p}$ in Eq.\ (12) implies that we only consider
the {\em incoherent} tunneling processes, which are known to be appropriate
for the defect-populated insulating barriers and the rough interfaces.
As usual, we make the following assumptions that the strengths of various
tunneling amplitudes, that is, $|T^{0}_{\bf kp}|^{2}, |T'_{\bf kp}|^{2}$,
and $|T''_{\bf kpq}|^{2}$ are independent of momenta ${\bf k,p,q}$ and then 
will be  taken to be their average values $|T^{0}|^{2}, |T'|^{2}$, 
and $|T''|^{2}$, respectively.
By making a transformation for the tunneling Hamiltonian from the collinear
magnetization configurations to
arbitrary relative angle $\theta$ between the spin orientations of the two
electrodes\cite{Lyu3} and using the standard Green's
function methods,\cite{mahan} one can obtain the tunneling conductance
$G(\theta)$ at zero bias up to the leading order in the tunneling amplitudes, 
which is given by

$$
G(\theta)=\frac{\pi e^{2}}{\hbar}|T^{0}|^{2}[\rho_{\downarrow}(\varepsilon_{F})+
\rho_{\uparrow}(\varepsilon_{F})]^{2}
[1+P^{2}{\rm cos}\theta $$
\begin{equation}
+\gamma(1-P^{2}{\rm cos}\theta)],
\end{equation}
where $P$ stands for the itinerant carrier spin polarization given by 
\begin{equation}
P=\frac{\rho_{\downarrow}(\varepsilon_{F})-\rho_{\uparrow}(\varepsilon_{F})}
{\rho_{\downarrow}(\varepsilon_{F})+\rho_{\uparrow}(\varepsilon_{F})},
\end{equation}
and $\rho_{\downarrow}(\varepsilon_{F})$ ($\rho_{\uparrow}(\varepsilon_{F})$)
is the density of states (DOS) of the spin-down (spin-up) carriers at the
Fermi level of the DMS electrodes. The relative angle $\theta$ may be
tunable by applying a magnetic field and $\gamma$ is given by the following
relation
\begin{equation}
\gamma=\gamma_{1}+4\eta cS(cS-\langle S^{z}\rangle),
\end{equation}
where
$\gamma _{1}= |T'|^{2}/|\widetilde{T}|^{2}$ and
$\eta = |T''|^{2}/|\widetilde{T}|^{2}$ with
$ |\widetilde{T}|^{2}=|T^{0}|^{2}+2c^{2}S^{2}|T''|^{2}\simeq |T^{0}|^{2}$,
since $c^2 |T''|^{2}$ is usually about two orders of magnitude smaller than
$|T^{0}|^{2}.$\cite{zhang}  
In deriving Eq.\ (15), we have used the following relation 
$V^{-1} \sum _{{\bf |q|}<q_{c}} \langle b^{\dag}_{\bf q}b_{\bf q} \rangle=cS-\langle
S^{z}\rangle$.
The corresponding tunneling resistance is given by $R(\theta)=1/G(\theta)$.

The TMR ratio is defined by the following formula
\begin{equation}
\left(\frac{\Delta R}{R_{\rm P}}\right)_{\rm max}=\frac{R_{\rm AP}-R_{\rm P}}
{R_{\rm P}},
\end{equation}
where $R_{\rm AP}$ and $R_{\rm P}$ represent the tunneling resistance in the
antiparallel and parallel spin alignments of the two
ferromagnetic electrodes respectively, and
the maximum TMR ratio for the DMS tunnel junctions can be obtained as
\begin{equation}
\left(\frac{\Delta R}{R_{\rm P}}\right)_{\rm
max}=\frac{2(1-\gamma)P^{2}}{1-P^{2}+
\gamma (1+P^{2})}.
\end{equation}

If the electrodes of the DMS tunnel junctions are made of the non-degenerate
DMSs, the tunneling conductance
will be proportional to the itinerant carrier density and the
TMR ratio will be more appropriately given by Eq.\ (17)
with the carrier spin polarization replaced by
$P=(n_{\downarrow}-n_{\uparrow})/(n_{\downarrow}+n_{\uparrow})$.
The different definitions of the carrier spin polarization in the TMR ratio
for the degenerate and non-degenerate DMS tunnel junctions are due to the
distinctly different transport properties in the corresponding DMSs.

\section{results and discussions}
In Fig.\ 1 and Fig.\ 2 are illustrated the temperature dependences
of
the impurity Mn ion magnetization of the DMS, the itinerant-carrier spin
polarization, and the maximum
TMR ratio for the DMS tunnel junctions at several different itinerant-carrier 
densities $c^{*}$
with the following choice of material parameters: $\gamma _{1}=0.692$, 
$\eta=0.005~ {\rm nm}^{6} 
$, and 
$m^{*}=0.5m_{e}$, $J_{pd}=0.15~ {\rm eV~nm}^{3}$ appropriate for the 
{\em bulk} Ga$_{1-x}$Mn$_{x}$As compounds \cite{ohno} and fixed impurity Mn 
ion density $c= 1~ {\rm nm}^{-3}$, where $m_{e}$ is the free electron mass.
Figure 1 (a) shows the impurity Mn ion magnetization $\langle S^z \rangle$ 
normalized by $c$. 
One can notice that the critical temperature $T_c$ decreases with the further 
increase of the itinerant carrier density $c^*$ from 
$c^{*}=0.1~{\rm nm}^{-3}$. 
{}From the solid curves of Fig.\ 1 (b) corresponding to $c^{*}=0.1~{\rm nm}^{-3}$, 
one can see the following features that at relatively low temperatures,
the spin-up carrier band is completely lifted from the spin-down carrier 
band
by the large exchange gap $\Delta (T)$. At the mean-field level, this leads 
to the complete
spin polarization of the itinerant 
carriers, {\em i.e.} $P=100 \%$, exhibiting the half-metallic feature.
In this case the TMR ratio is reduced to $(\Delta R/R_{p})_{\rm max}
=1/ \gamma-1$, which
decreases with the increase in the population of the thermally excited spin 
wave collective excitations leading to the increase of $\gamma$ as shown in
Fig.\ 1 (c). As the temperature 
increases, the size of the exchange gap $\Delta (T)$ is
gradually reduced. With the further increase of temperature beyond a certain
value, the number of the majority spin-down carriers begins to decrease, 
while that of the minority spin-up carriers starts to increase.
At high temperatures, the carrier spin polarization
decreases rapidly with temperatures, which makes the
TMR effect disappear above a critical temperature. 
The factor of $(1- \gamma)$ in Eq. (17) indicates that the TMR effect may   
vanish below the critical temperature, which is due to the competition between 
the spin-conserving tunneling amplitude $|T^{0}|^{2}$ and the spin-flip 
tunneling ones $|T'|^{2}, |T''|^{2}$.

For higher carrier densities, say, $c^{*}=0.2~{\rm nm}^{-3}$,
the Fermi energy of the system increases and then both spin states are 
occupied losing the half-metallic feature as shown in Fig. 1 (b).
Since the partial carrier spin polarization is not
beneficial to the TMR effect, it leads to the low TMR ratio.
Figure 2 (a) illustrates the impurity Mn ion magnetization $\langle S^z \rangle$ 
normalized by $c$ at different values of $c^*$. In contrary to Fig. 1 (a), one can 
see that $T_c$ decreases with the  
decrease of the itinerant carrier density $c^*$ from 
$c^{*}=0.05~{\rm nm}^{-3}$ exhibiting the {\em non-monotonic} dependence of 
$T_c$ on $c^*$, which has been emphasized in Ref. \onlinecite{Mac}. 
For $c=1~{\rm nm}^{-3}$, the optimal carrier densities, which will achieve 
the maximum TMR ratio, are shown to be in the range 
of $0.05~{\rm nm}^{-3}<c^{*}<0.1~{\rm nm}^{-3}$.

Recently Hayashi {\em et al.} have performed an experiment for  
Ga$_{0.961}$Mn$_{0.039}$As/AlAs/Ga$_{0.961}$Mn$_{0.039}$As tunnel junctions and 
obtained the TMR ratio of about 44\% at 4.2 K.\cite{1}  
By fitting the above experimental data to our calculations, we were able to 
estimate the approximate values of material parameters: 
$c^{*}=0.13~{\rm nm}^{-3}$, $c=0.86~{\rm nm}^{-3}$, and $\gamma _{1}=0.692$. 
The above value of carrier density $c^{*}$ is about 15 $\%$ of the Mn
ion concentration $c$.
The value of $\gamma _{1}$ indicates that there exists strong spin-flip
scattering by the Mn ions within the insulating barrier. 
During the fabrication process of the DMS
tunnel junctions, the highly doped Mn ions in the Ga$_{1-x}$Mn$_{x}$As layers
can easily diffuse into the AlAs insulating layer. Reducing the Mn ions in the
insulating layer is crucial to achieve a large TMR ratio in the DMS tunnel
junctions.

Figure 3 shows the temperature dependences of the impurity Mn
ion magnetization of the DMS, the itinerant-carrier spin polarization,
and the maximum TMR ratio for the DMS tunnel junctions at several values of 
impurity Mn ion
densities $c$ with fixed carrier density $c^{*}=0.1~{\rm nm}^{-3}$. Figures 3 (a) 
and (b) demonstrate that the carrier band structure is mainly determined 
from the size of the exchange gap 
$\Delta (T)$. At low impurity densities, say, $c=0.6~{\rm nm}^{-3}$, 
the strength of the ferromagnetic coupling is weak, which induces partial 
carrier spin polarization. 
By increasing $c$, the ferromagnetic coupling becomes strong, which completely 
polarizes the carrier spins above a certain value of $c$.
So achieving the high doping of the impurity Mn ions is
necessary to have much higher $T_c$ and larger TMR ratio at high temperatures.

When the magnetic ion density is beyond a certain large value, say, $c= 1.5~{\rm nm}^{-3}$, 
the impurity ion
magnetization and the itinerant-carrier spin polarization seem to exhibit
the behaviors of the first order ferromagnetic transition as shown in Fig. 3 (a) and (b).  
In this case, the first order ferromagnetic transition is induced by 
the $p$-$d$ exchange coupling,
in contrast to that for the two-dimensional quantum well system 
at very low carrier densities, which is argued to be due to the 
carrier-carrier correlation effects.\cite {brey} 

Our results clearly demonstrate that the TMR effects in the DMS tunnel 
junctions are dominated
by both the ferromagnetic coupling strength and the carrier band structure through the
impurity ion density and the itinerant carrier density in the DMS electrodes.
The elastic and inelastic spin-flip tunnelings play an important role in
reducing the TMR ratio.  

\section{summary}
In summary, we have presented the TMR formula for the DMS tunnel junctions by generalizing
the previous spin-polarized tunneling model and taking into account
the basic physics of the DMS.
The experimentally observed TMR ratio was reproduced
by the appropriate choice of the typical material parameters. It was also shown that
the TMR ratio has a strong dependence on the
itinerant-carrier density and the magnetic ion density in the DMS electrodes 
as well.
In view of the recent predictions of the room
temperature ferromagnetic semiconductors,\cite{4} the DMS tunnel junctions 
can be practically very useful in the future digital storage and magnetic 
sensor technology.
Our analysis of the TMR ratio can be potentially useful to 
achieve larger TMR ratio by optimally adjusting the material parameters.

\section{acknowledgments}
This work was supported by the Brain Korea 21 Project and also by Grant No.
1999-2-112-001-5 from the interdisciplinary Research program of the KOSEF.



\vspace*{15cm}
\includegraphics{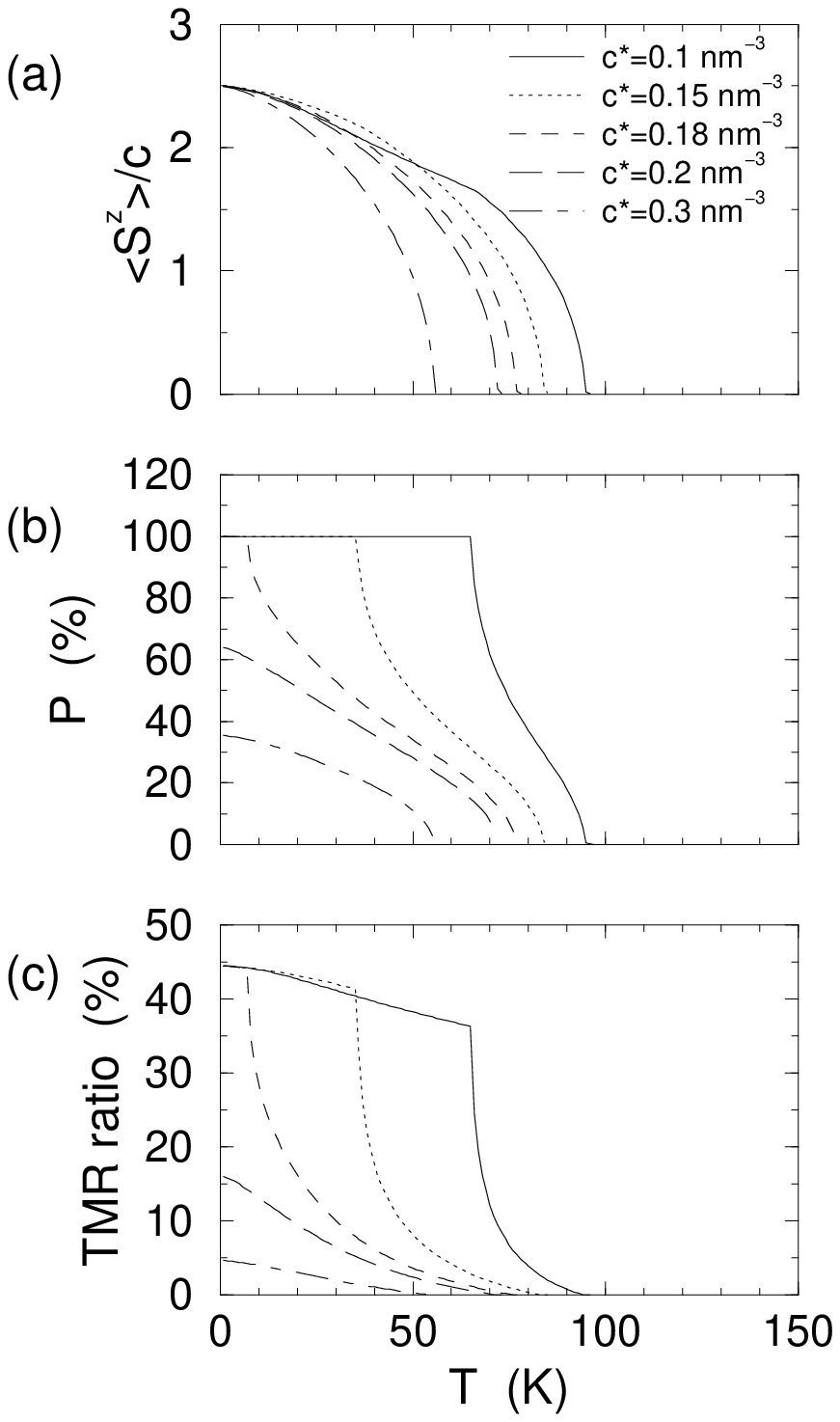}

Fig.\ 1.\ (a) Impurity Mn ion magnetization of the DMS, (b) itinerant-carrier
spin polarization $P$, and (c) the maximum TMR ratio in the DMS tunnel 
junctions as a function of temperature for several different 
itinerant-carrier densities $c^{*}=0.3,~0.2,~0.18,~0.15,~0.1~{\rm nm}^{-3}$
with the following choice of material parameters $m^{*}=0.5m_{e}$, 
$J_{pd}=0.15~ {\rm eV~nm}^{3}$, $\gamma _{1}=0.692$ and 
$\eta=0.005~{\rm nm}^{6}$, and the fixed value of $c= 1~ {\rm nm}^{-3}$.

\vspace*{17cm}

\includegraphics{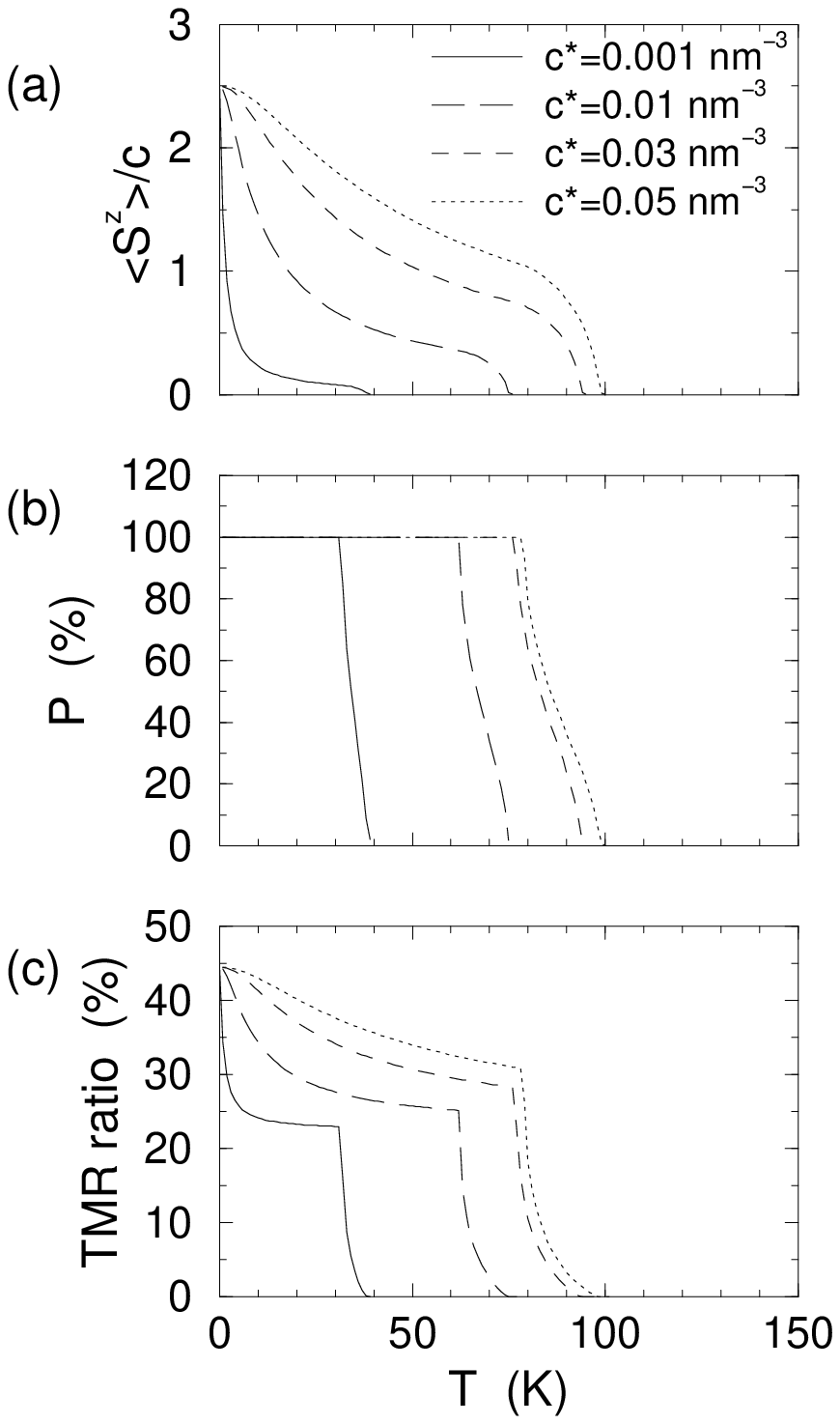}

Fig.\ 2.\ (a) Impurity Mn ion magnetization of the DMS, (b) itinerant-carrier
spin polarization $P$, and (c) the maximum TMR ratio in the DMS tunnel 
junctions as a function of temperature at several different itinerant-carrier 
densities $c^{*}=0.05,~ 0.03,~ 0.01,~0.001 ~{\rm nm}^{-3}$
with the following choice of material parameters $m^{*}=0.5m_{e}$, 
$J_{pd}=0.15~ {\rm eV~nm}^{3}$,
 $\gamma _{1}=0.692$ and $\eta=0.005~{\rm nm}^{6}$, and the fixed 
value of $c= 1~ {\rm nm}^{-3}$.

\vspace*{17cm}

\includegraphics{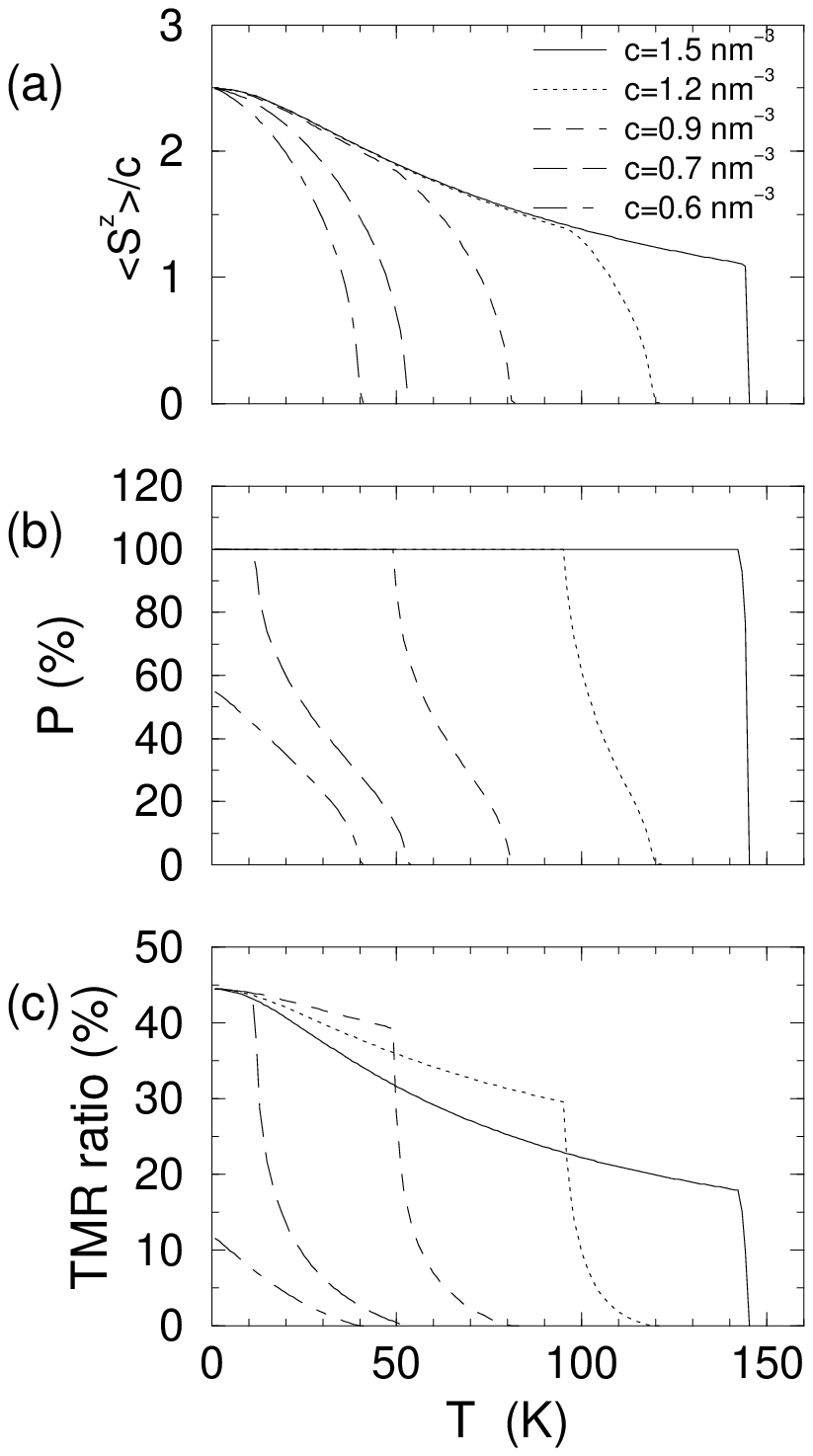}

Fig.\ 3.\ (a) Impurity Mn ion magnetization of the DMS, (b) itinerant-carrier
spin polarization $P$, and (c) the maximum TMR ratio in the DMS tunnel 
junctions as a function of temperature for several different impurity Mn ion 
densities
$c=0.6$, 0.7, 0.9, 1.2, 1.5 ${\rm nm}^{-3}$
with the following choice of material parameters $m^{*}=0.5m_{e}$, 
$J_{pd}=0.15~ {\rm eV~nm}^{3}$,
 $\gamma _{1}=0.692$ and $\eta=0.005~{\rm nm}^{6}$, and the fixed 
value of $c^{*}=0.1~ {\rm nm}^{-3}$.

\end{document}